\documentclass
[prl,twocolumn,superscriptaddress,floatfix,showpacs,10pt]{revtex4}%
\usepackage{amsfonts}
\usepackage{amsmath}
\usepackage{amssymb}
\usepackage{graphicx}%
\begin{document}
\title{Thermodynamics of \textquotedblleft exotic\textquotedblright\ BTZ black holes}
\author{Paul K. Townsend}
\email{p.k.townsend@damtp.cam.ac.uk}
\affiliation{Department of Applied Mathematics and Theoretical Physics, Centre for
Mathematical Sciences, University of Cambridge, Wilberforce Road, Cambridge,
CB3 0WA, UK}
\author{Baocheng Zhang}
\email{zhangbc@wipm.ac.cn}
\affiliation{State Key Laboratory of Magnetic Resonances and Atomic and Molecular Physics,
Wuhan Institute of Physics and Mathematics, Chinese Academy of Sciences, Wuhan
430071, China}

\begin{titlepage}
\begin{flushright}  DAMTP-2013-9
\end{flushright}
\vfill
\end{titlepage}

\begin{abstract}

A number of  three-dimensional (3D) gravity models, such as 3D conformal gravity, admit ``exotic'' black 
hole solutions: the metric is the same as the BTZ metric of 3D Einstein gravity but with reversed  roles for 
mass and angular momentum, and an entropy proportional to the length  of the {\it inner}  horizon instead 
of the event horizon. Here we show that the BTZ solutions of the ``exotic'' 3D Einstein gravity (with 
parity-odd action but Einstein field equations) are exotic black holes, and we investigate their thermodynamics.
The first and second laws of black hole thermodynamics  still apply, and the entropy still has a statistical interpretation.

\end{abstract}

\maketitle

\setcounter{equation}{0}

In three spacetime dimensions (3D) the Einstein gravitational field equations
imply that the metric has constant curvature and hence, in the case of a
negative cosmological constant, that it is locally isometric to anti-de Sitter
(adS) space. One solution is therefore the adS vacuum but there is also a
two-parameter family of stationary black hole solutions, found by Banados,
Teitelboim and Zanelli  \cite{Banados:1992wn}; the BTZ metric takes the form 
\begin{equation}
ds^{2}=-N^{2}dt^{2}+N^{-2}dr^{2}+r^{2}\left(  d\phi+N^{\phi}dt\right)  ^{2}\,. 
\label{btz}%
\end{equation}
The functions $N^2$ and $N^\phi$ are 
\begin{equation}
N^{2}=-8Gm+\frac{r^{2}}{\ell^{2}}+\frac{16G^{2}j^{2}}{r^{2}}\,,\qquad N^{\phi
}=-\frac{4Gj}{r^{2}}\,, 
\end{equation}
where $\ell$ is the adS radius and $G$ the 3D Newton constant. 
The parameters $(m,j)$ can be interpreted as the mass $M$ and angular 
momentum $J$ of the black hole in units for which $\hbar=1$; i.e. $M=m$ and 
$J=j$. The laws of black hole thermodynamics \cite{Bardeen:1973gs} apply, with the entropy
given by the Bekenstein-Hawking formula \cite{Bekenstein:1973ur,Hawking:1974sw}, which 
states (in the 3D context) that the entropy is 1/4 of the length of the event horizon in 
Planck units. There is even a statistical mechanics interpretation of this entropy
\cite{Strominger:1997eq,Birmingham:1998jt};  the microstates are those of a holographically dual
conformal field theory (CFT)  implicit in earlier work of Brown and Henneaux  \cite{Brown:1986nw}, who 
showed that the central charges are
\begin{equation}
c_{L}=c_{R}=\frac{3\ell}{2G}\,. \label{brown-henneaux}%
\end{equation}
The gravity description of the CFT is valid at large $\ell/G$.

Because the BTZ metric is locally isometric to adS$_{3}$, it solves the field
equations of \textit{any} 3D gravity model admitting an adS$_3$ vacuum, and the
mass and angular momentum may then be equal to some other linear combination
of the parameters. Rather surprisingly, a number of 3D gravity models have
been found \cite{Carlip:1991zk,Carlip:1994hq,Banados:1997df,Banados:1998dc}
for which the roles of mass and angular momentum are \textit{reversed}, in the
sense that
\begin{equation}
M=j/\ell\,,\qquad J=\ell m\,.
\end{equation}
Such BTZ black holes have been called \textquotedblleft
exotic\textquotedblright. The latest addition \cite{Afshar:2011qw} to the
class of 3D gravity models for which BTZ black holes are exotic is 3D
conformal gravity \cite{VanNieuwenhuizen:1985ff,Horne:1988jf}, for which the
simplest action is the Lorentz-Chern-Simons (LCS) term,  introduced in 
\cite{Deser:1981wh} in the context of  \textquotedblleft topologically
massive gravity\textquotedblright. The central charges of the CFT
holographically dual to generic 3D gravity models may be computed by a
variety of methods \cite{Kraus:2005zm,Afshar:2011qw}, and for models admitting
exotic BTZ black holes it is found, e.g. for conformal 3D gravity, that
\begin{equation}
c_{R}=-c_{L}\,.
\end{equation}
This is what  one might expect of a gravity model with a parity-odd action.

Like the Kerr solution of the 4D Einstein equations, the BTZ metric has two
Killing horizons, located  at $r=r_\pm$, where $r_\pm$ are the zeros of $N^2$: 
\begin{equation}
r_{\pm}=\sqrt{2G\ell\left(  \ell m+j\right)  }\pm\sqrt{2G\ell\left(  \ell
m-j\right)  }\,. \label{KH}%
\end{equation}
We may assume without loss of generality that $j\geq0$. We shall also assume
that $\ell m\geq j$, to ensure the existence of an event horizon at $r=r_+$. 
Note that the parameters $(m,j)$ can be simply expressed in terms of the radii of the
inner and outer horizons:
\begin{equation}
\ell m=\frac{r_{+}^{2}+r_{-}^{2}}{8G\ell}\,,\qquad j=\frac{r_{+}r_{-}}{4G\ell
}\,. \label{jandm}%
\end{equation}

The entropy of BTZ black holes has also been computed by  various means, which include 
holographic methods \cite{Solodukhin:2005ah} and  Wald's Noether charge formula \cite{Wald:1993nt}, 
or its extension  to allow for parity      violation \cite{Tachikawa:2006sz}.  Applied to the original BTZ black hole,
these methods give a result in agreement with the Bekenstein-Hawking formula, but applied to
exotic BTZ black holes they give a different result: the entropy is proportional to the length $2\pi r_-$ of the \textit{inner} horizon.

This result calls into question the validity of the laws of black hole
thermodynamics. For example, a perturbation of a ``normal'' BTZ black hole
will \textit{decrease} the length of its inner horizon
\cite{Park:2006pb,Detournay:2012ug}. This is not problematic in itself but it
illustrates the point that Hawking's area theorem (length in 3D) need not
apply to inner horizons, and as the entropy of an exotic BTZ black hole is
proportional to the length of the inner horizon, it becomes unclear whether
the second law of thermodynamics still applies. In an attempt to resolve some
of these difficulties, Park proposed a new formula for the BTZ black hole
entropy in the context of higher-derivative 3D gravity theories
\cite{Park:2006hu}. His proposal was supported by a computation of the
statistical entropy but it conflicts with results obtained by other methods
and implies various thermodynamic abnormalities, such as negative temperature.

Here we investigate these issues in the context of what must surely be the
simplest 3D gravity model for which BTZ black holes are exotic. This is the
\textit{parity-odd}  action for 3D Einstein gravity with negative cosmological constant
that Witten called (coincidentally)   \textquotedblleft exotic\textquotedblright  \cite{Witten:1988hc}. 
To contrast the normal and exotic versions of 3D Einstein gravity, we give
here their respective Lagrangian 3-forms in terms of independent dreibein
1-forms $e^{a}$ and Lorentz connection 1-forms $\omega^{a}$ ($a=0,1,2$), and
their torsion and curvature 2-form field strengths,
\begin{equation}
T^{a}=de^{a}+\epsilon^{abc}\omega_{b}e_{c}\,,\qquad R^{a}=d\omega^{a}+\frac
{1}{2}\epsilon^{abc}\omega_{b}\omega_{c}\,,
\end{equation}
where the exterior product of forms is implicit. In the normal case the
Lagrangian 3-form is
\begin{equation}
L=\frac{1}{8\pi G}\left[  e_{a}R^{a}-\frac{1}{6\ell^{2}}\epsilon^{abc}%
e_{a}e_{b}e_{c}\right]  \,. \label{3fN}%
\end{equation}
This is the well-known Einstein-Cartan formulation, so we do not elaborate
further. The Lagrangian 3-form for the exotic theory is
\begin{equation}
L_{E}=\frac{\ell}{8\pi G}\left[  \omega_{a}\left(  d\omega^{a}+\frac{2}%
{3}\epsilon^{abc}\omega_{b}\omega_{c}\right)  -\frac{1}{\ell^{2}}e_{a}%
T^{a}\right]  \,. \label{3fE}%
\end{equation}
The first term is the LCS term (except that the connection is an independent
one). Although the action obtained by integration of $L_{E}$ does not preserve
parity, its parity transform is minus itself (i.e. it is parity-odd) so the
field equations do preserve parity. Varying with respect to $e^{a}$, we get
the torsion constraint $T^{a}=0$, which allows us to solve for $\omega^{a}$;
using this in the equation obtained by varying $\omega^a$,  we recover the same
Einstein equations that follow from the standard  parity-even Lagrangian
3-form (\ref{3fN}).

An explanation of how the exotic action was found will also explain why its
BTZ black holes are exotic. The 3D adS isometry group is $SO(2,2)\cong Sl(2;%
\mathbb{R}
)\times Sl(2;%
\mathbb{R}
)$. It was shown in \cite{Achucarro:1987vz} that the field equations of the
$SO(2,2)$ Chern-Simons (CS) action are equivalent to the Einstein equations
with negative cosmological constant. Actually, since one may consider any
linear combination of the two CS terms for the $Sl(2;%
\mathbb{R}
)$ factors, for which the 1-form potentials are
\begin{equation}
A_{\pm}^{a}=e^{a}\pm\ell\omega^{a}\,, \label{Apm}%
\end{equation}
the choice of $SO(2,2)$ Chern-Simons term is not unique, although there is a
unique choice that yields a parity-preserving action. Because CS terms are
intrinsically parity odd, and because the 1-forms $A_{+}^{a}$ and $A_{-}^{a}$
are interchanged by parity, only the \textit{difference} of the two $Sl(2;\mathbb{R})$ CS terms
preserves parity, and for a suitable overall constant this combination yields
the standard action of 3D general relativity. This observation was first made
by Ach\'{u}carro \cite{MSC} and later by Witten \cite{Witten:1988hc}, who
also wrote down an explicit form for the  parity-odd action obtained by taking the sum of
the two $Sl(2;%
\mathbb{R}
)$ CS terms.  Written in terms of the 1-forms
$(e^{a},\omega^{a})$ this ``exotic''  version has the Lagrangian 3-form of
(\ref{3fE}).

Let us now examine the implications for BTZ black holes. For each of the two
$Sl(2;%
\mathbb{R}
)$ subgroups of $SO(2,2)$, we can define a conserved charge $Q_{\pm}$ as the
holonomy of an asymptotic $U(1)$ connection \cite{Izquierdo:1994jz}. Using
(\ref{Apm}) one then finds, in the \textquotedblleft normal\textquotedblright%
\ case, that
\begin{equation}
\ell M=Q_{+}+Q_{-}\,,\qquad J=Q_{+}-Q_{-}\,,\qquad(\mathrm{normal})
\end{equation}
so parity, which exchanges $Q_\pm$,  leaves $M$ unchanged but flips the sign of $J$, as expected. In
contrast, the exotic case gives
\begin{equation}
\ell M=Q_{+}-Q_{-}\,,\qquad J=Q_{+}+Q_{-}\,,\qquad(\mathrm{exotic})
\end{equation}
so the roles of $\ell M$ and $J$ are reversed. For the BTZ solution of
\textquotedblleft normal\textquotedblright\ 3D gravity we have $M=m$ and
$J=j$, so we will find that $\ell M=j$ and $J=\ell m$ when we view it as a
solution of the \textquotedblleft exotic\textquotedblright\ theory. In other
words, \textit{BTZ black holes are \textquotedblleft normal\textquotedblright%
\ as solutions of \textquotedblleft normal' 3D Einstein gravity, and
\textquotedblleft exotic\textquotedblright\ as solutions of \textquotedblleft
exotic\textquotedblright\ 3D Einstein gravity}.

We now turn to black hole thermodynamics.  
The Hawking temperature $T$ of the BTZ black hole and the angular velocity
$\Omega$ of its event horizon are
\begin{equation}
T=\frac{r_{+}^{2}-r_{-}^{2}}{2\pi r_{+}\ell^{2}}\,,\qquad\Omega=\frac{r_{-}%
}{\ell r_{+}}\,. \label{bht}%
\end{equation}
These expressions for the intensive thermodynamic variables are geometrical in the sense that they depend only on the
location of the Killing horizons and are model-independent, i.e. independent
of the particular field equations that are solved by the BTZ metric.

In contrast,  the extensive thermodynamic variables are model-dependent. For generality, we shall 
consider the case in which the mass and angular momentum are given by
\begin{equation}
M=\alpha m+\gamma j/\ell\,,\qquad J=\alpha j+\gamma\ell m\,,\label{MandJ}%
\end{equation}
for constants $(\alpha,\gamma)$. The cases $(\alpha,\gamma)=(1,0)$ and $(\alpha,\gamma)=(0,1)$ correspond,
respectively, to the normal and exotic BTZ black holes.  In the ``normal'' case the second of the 
relations (\ref{jandm}) tells us that the product of the lengths of the inner and outer horizons is independent of $M$, 
as  expected on general grounds \cite{Larsen:1997ge,Cvetic:2010mn,Castro:2012av,Ansorg:2009yi}, but 
 in the ``exotic'' case it is independent of $J$. 

Let us now write the first law of black hole thermodynamics as
\begin{equation}
dM-\Omega dJ=TdS\, .\label{bh1}%
\end{equation}
We claim that this law is satisfied for $M$ and $J$ given by (\ref{MandJ}) if, and only if, 
the entropy $S$ is given by
\begin{equation}\label{genentropy}
S=\frac{\pi}{2G}\left(  \alpha r_{+}+\gamma r_{-}\right)  \,.
\end{equation}
In fact, given only $M$ as a linear combination of $m$ and $j$, the expressions for both $J$ and $S$ can be fixed by requiring 
the validity of the first law.  This can be verified by computing $dS$ in terms of $dM$ and $dJ$ using (\ref{KH}) and (\ref{bht}).
When  $(\alpha,\gamma)=(1,0)$ the entropy reduces to the usual Bekenstein-Hawking formula, but when  $(\alpha
,\gamma)=(0,1)$ it is given by the ``exotic'' formula $S=S_E\equiv \pi r_-/(2G)$; i.e. it is now $1/4$  the length,  in Planck units, 
 of the {\it inner}  horizon. 

Next we show that the entropy, as given by (\ref{genentropy}), obeys the second law, at least for the  quasi-stationary process of
infalling matter. Let $p$ be the $3$-momentum of a particle falling through
the event horizon. Since the event horizon is a Killing horizon for the
Killing vector
\begin{equation}
\xi= \partial_{t} + \Omega\partial_{\phi}\, ,
\end{equation}
we have $-p\cdot\xi\ge0$ at the horizon. This implies that the changes $dM$
and $dJ$ in the mass and angular momentum of the black hole satisfy the
inequality $dM \ge\Omega dJ$, and hence that
\begin{equation}
\ell r_{+} dM \ge r_{-} dJ\, .
\end{equation}
Using the expressions (\ref{MandJ}) for the mass and angular momentum, and the
formulas (\ref{KH}) for $r_{\pm}$, we find that
\begin{equation}
dS \ge0\, .
\end{equation}
This is true for any constants $(\alpha,\gamma)$, in particular it is true for the
exotic case $(\alpha,\gamma)=(0,1)$, even though  the entropy is then proportional to the 
length of the inner horizon!

It should be noted that, in arriving at these results, we  have used the angular momentum and temperature
associated to the \textit{event} horizon, not to the inner horizon. We are
\textit{not} discussing here the \textquotedblleft thermodynamics of inner
horizons\textquotedblright. 

Finally, we consider  the statistical entropy of exotic BTZ black holes. In the
high-temperature ($\beta\rightarrow0$) limit, the partition function
$Z(\beta)=\mathrm{tr}\,e^{-\beta H}$ of a CFT with Hamiltonian $H$ can be
approximated by the integral
\begin{equation}
Z(\beta)=\int_{0}^{\infty}\!\!d\Delta\,\rho(\Delta)e^{-\beta\Delta}\,,
\end{equation}
where $\Delta$ is the eigenvalue of the Virasoro generator $L_{0}$, and
$\rho(\Delta)=e^{S(\Delta)}$ is the (smoothed) density of states, equal to the
exponential of the entropy function $S(\Delta)$. Using the fact that
$S\propto\sqrt{\Delta}$, the partition function can be evaluated in a saddle
point approximation. Then, using modular invariance, the result can be related
to the low temperature ($\beta\rightarrow\infty$) limit, which is dominated by
the ground state, with energy determined by the central charge. In this way
Cardy found a large-$\Delta$ approximation to $\rho(\Delta)$, such that
\cite{Cardy:1986ie}
\begin{equation}
S(\Delta)\approx2\pi\sqrt{c\Delta/6}\,, \label{Cardy}%
\end{equation}
where $c$ is the central charge. Applying this formula to the left and right
sectors of the Brown-Henneaux CFT with central charges $c_{L}=c_{R}%
=3\ell/(2G)$, and using the relations
\begin{equation}
2\Delta_{L}=\ell M+J\,,\qquad2\Delta_{R}=\ell M-J\,, \label{Deltas}%
\end{equation}
one finds that \cite{Strominger:1997eq,Birmingham:1998jt}
\begin{equation}
S_{L}=\pi\sqrt{\frac{\ell\left(  \ell M+J\right)  }{2G}}\,,\qquad S_{R}%
=\pi\sqrt{\frac{\ell\left(  \ell M-J\right)  }{2G}}\,.
\end{equation}
The sum $S=S_{L}+S_{R}$ equals the Bekenstein-Hawking entropy $\pi r_{+}/(2G)$.
We may also conclude from this result that the partition function of the Brown-Henneaux
CFT takes the holomorphically factorized form $Z= Z_LZ_R$ within the limits of the 
leading-order saddle-point approximation. Whether this form holds exactly is not known; we defer 
to \cite{Maloney:2007ud} for a discussion of this point. 

In the exotic case we expect to have $c_{R}=-c_{L}$, as in the 3D conformal
gravity case, because the action is parity odd. This means that we expect
\begin{equation}
c_{L} = \frac{3\ell}{2G}\, , \qquad c_{R} = -\frac{3\ell}{2G}\, .
\end{equation}
We might also expect to have $2\Delta_{L}= \ell M_{E} +J_{E}$ and $2\Delta_{R}
= J_{E}-\ell M_{E}$ because this is what one gets from (\ref{Deltas}) by
setting $J= \ell M_{E}$ and $\ell M = J_{E}$. The weak cosmic censorship condition
$J_{E}\ge\ell M_{E}$, needed for the existence of a horizon, then ensures that
$\Delta_{R}>0$. However, if this were true then we would have $c_{R}\Delta
_{R}<0$ and the Cardy formula would give an imaginary entropy.

Given that $c_{R}<0$, the ``exotic'' version of the Brown-Henneaux CFT must be
non-unitary, but this is only because we consider both the left and right
sectors together. We may consider each sector in isolation because there is no
interaction between them (at least within the approximation reviewed above). Taken in isolation, there is no significance to the
sign of the central charge $c_{R}$ because we can change its sign from
positive to negative by declaring that physical states have negative norm
instead of positive norm, so that the energy is now bounded from above rather
than from below. This is just a change of conventions; only the sign of
$c_{R}$ relative to $c_{L}$ is convention-independent. We suggest that
$c_{R}<0$ for exotic 3D gravity theories precisely because the conventions for
the right-movers are the reverse of the usual ones, in the sense just
explained. This would flip the sign of $\Delta_{R}$ so that
\begin{equation}
2\Delta_{L}= \ell M_{E} +J_{E}\, , \qquad-2 \Delta_{R} = J_{E}-\ell M_{E}\, .
\end{equation}
Observe that the weak cosmic censorship condition $J_{E}-\ell M_{E}$ now
ensures that $\Delta_{R}<0$, and hence $c_{R}\Delta_{R}>0$. Applying the Cardy
formula (\ref{Cardy}) we find that
\begin{equation}
S_{L}= \pi\sqrt{\frac{\ell\left(  J+ \ell M\right)  }{2G}}\, , \qquad S_{R}=
\pi\sqrt{\frac{\ell\left(  J-\ell M\right)  }{2G}}\, .
\end{equation}

Now we must ask how the partition function $Z_{R}$ changes if we change the
sign of the norm of all physical states so that $c_{R}\to-c_{R}$ and all
energies change sign. We have seen  that the laws of black hole 
thermodynamics continue to apply to exotic BTZ black holes, so we expect
standard thermodynamic relations such as $E=-\partial\ln Z/\partial\beta$ to
remain valid, but this relation is maintained when $E\to-E$ only if we also take
$Z\to Z^{-1}$. We conclude from this that whereas $Z=Z_{L} Z_{R}$ in the
``normal'' case, we must have $Z=Z_{L}/Z_{R}$ in the exotic case, and this
means that we should now \textit{subtract} $S_{R}$ from $S_{L}$ to get the
total ``exotic'' entropy: 
\begin{equation}
S_{E} = \pi\sqrt{\frac{\ell\left(  J+ \ell M\right)  }{2G}} - \pi\sqrt
{\frac{\ell\left(  J-\ell M\right)  }{2G}} = \frac{\pi r_{-}}{2G}\, . 
\end{equation}
This is precisely the entropy required for  the validity of the laws of black hole thermodynamics 
for exotic BTZ black holes.  As far as we are aware, this is the first time that  a statistical
interpretation has been found for a black hole entropy that is {\it not} given by the Bekenstein-Hawking formula. 

In this paper, we have addressed puzzles that have
arisen in the study of ``exotic''  BTZ black holes, for which
the three-dimensional spacetime is that of the usual BTZ black hole but with a
reversal of the roles of mass and angular momentum.  We have shown that such
 black holes occur even for 3D Einstein gravity in the context of its parity-odd 
 ``exotic'' action.  The relationship of the ``normal'' to the ``exotic'' versions of 3D 
Einstein gravity explains the reversal of the roles of mass and angular momentum. 
It also  goes a long way to explaining why the BTZ black holes of conformal 3D gravity are 
exotic: it is because the exotic Einstein
gravity can be viewed as a truncation of conformal 3D gravity; from the
Chern-Simons perspective one is just restricting to an $SO(2,2)$ subgroup of
$SO(2,3)$.

Our  investigation of  the thermodynamics of BTZ black holes allowed  for the mass
and angular momentum to be given by arbitrary linear combinations of the parameters of the 
BTZ metric. In fact,  given only the mass,  the first law determines the entropy, which then satisfies
the second law, at least  for quasi-static perturbations, and this is a {\it general result} applicable 
to all 3D gravity models. For exotic black holes the entropy is  equal to  $1/4$ of the length of the {\it inner} 
horizon, and we have presented a statistical interpretation of this (non-Bekenstein-Hawking) result in terms of a dual CFT, using 
holomorphic factorization of the leading order partition function  for 3D Einstein gravity. It  is not known whether
the exact partition function factorizes, but  if it does not there will be non-holomorphic interactions  that 
will almost certainly imply instabilities if  $c_R=-c_L$, so holomorphic factorizability may be an essential requirement 
for any 3D quantum gravity for which the BTZ black holes are exotic.

\noindent\textbf{Acknowledgements}: PKT thanks Gary Gibbons and David Tong for
helpful conversations. BZ is grateful for the hospitality of the General
Relativity group in the Department of Applied Mathematics and Theoretical
Physics of the University of Cambridge, where this work was begun, and he also
acknowledges support from Grant No. 11104324 of the National Natural Science
Foundation of China.

\end{document}